\documentclass[%
 reprint,
superscriptaddress,
 amsmath,amssymb,
 aps,
 prl,
 letterpaper,
 raggedbottom,
 nobalancelastpage]{revtex4-1}

\usepackage{graphicx}
\usepackage{dcolumn}
\usepackage{bm}

\begin{document}

\title{Acceleration of electrons in the plasma wakefield of a proton bunch}
\author{E.~Adli}
\affiliation{University of Oslo, Oslo, Norway}
\author{A.~Ahuja}
\affiliation{CERN, Geneva, Switzerland}
\author{O.~Apsimon}
\affiliation{University of Manchester, Manchester, UK}
\affiliation{Cockcroft Institute, Daresbury, UK}
\author{R.~Apsimon}
\affiliation{Lancaster University, Lancaster, UK}
\affiliation{Cockcroft Institute, Daresbury, UK}
\author{A.-M.~Bachmann}
\affiliation{CERN, Geneva, Switzerland}
\affiliation{Max Planck Institute for Physics, Munich, Germany}
\affiliation{Technical University Munich, Munich, Germany}
\author{D.~Barrientos}
\affiliation{CERN, Geneva, Switzerland}
\author{F.~Batsch}
\affiliation{CERN, Geneva, Switzerland}
\affiliation{Max Planck Institute for Physics, Munich, Germany}
\affiliation{Technical University Munich, Munich, Germany}
\author{J.~Bauche}
\affiliation{CERN, Geneva, Switzerland}
\author{V.K.~Berglyd Olsen}
\affiliation{University of Oslo, Oslo, Norway}
\author{M.~Bernardini}
\affiliation{CERN, Geneva, Switzerland}
\author{T.~Bohl}
\affiliation{CERN, Geneva, Switzerland}
\author{C.~Bracco}
\affiliation{CERN, Geneva, Switzerland}
\author{F.~Braunm{\"u}ller}
\affiliation{Max Planck Institute for Physics, Munich, Germany}
\author{G.~Burt}
\affiliation{Lancaster University, Lancaster, UK}
\affiliation{Cockcroft Institute, Daresbury, UK}
\author{B.~Buttensch{\"o}n}
\affiliation{Max Planck Institute for Plasma Physics, Greifswald, Germany}
\author{A.~Caldwell}
\affiliation{Max Planck Institute for Physics, Munich, Germany}
\author{M.~Cascella}
\affiliation{UCL, London, UK}
\author{ J.~Chappell}
\affiliation{UCL, London, UK}
\author{E.~Chevallay}
\affiliation{CERN, Geneva, Switzerland}
\author{M.~Chung}
\affiliation{UNIST, Ulsan, Republic of Korea}
\author{D.~Cooke}
\affiliation{UCL, London, UK}
\author{H.~Damerau}
\affiliation{CERN, Geneva, Switzerland}
\author{L.~Deacon}
\affiliation{UCL, London, UK}
\author{L.H.~Deubner}
\affiliation{Philipps-Universit{\"a}t Marburg, Marburg, Germany}
\author{A.~Dexter}
\affiliation{Lancaster University, Lancaster, UK}
\affiliation{Cockcroft Institute, Daresbury, UK}
\author{S.~Doebert}
\affiliation{CERN, Geneva, Switzerland}
\author{J.~Farmer}
\affiliation{Heinrich-Heine-University of D{\"u}sseldorf, D{\"u}sseldorf, Germany}
\author{V.N.~Fedosseev}
\affiliation{CERN, Geneva, Switzerland}
\author{R.~Fiorito}
\affiliation{University of Liverpool, Liverpool, UK}
\affiliation{Cockcroft Institute, Daresbury, UK}
\author{R.A.~Fonseca}
\affiliation{ISCTE - Instituto Universit\'{e}ario de Lisboa, Portugal}
\author{F.~Friebel}
\affiliation{CERN, Geneva, Switzerland}
\author{L.~Garolfi}
\affiliation{CERN, Geneva, Switzerland}
\author{S.~Gessner}
\affiliation{CERN, Geneva, Switzerland} 
\author{I.~Gorgisyan}
\affiliation{CERN, Geneva, Switzerland}
\author{A.A.~Gorn}
\affiliation{Budker Institute of Nuclear Physics SB RAS, Novosibirsk, Russia} 
\affiliation{Novosibirsk State University, Novosibirsk, Russia}
\author{E.~Granados}
\affiliation{CERN, Geneva, Switzerland}
\author{O.~Grulke}
\affiliation{Max Planck Institute for Plasma Physics, Greifswald, Germany}
\affiliation{Technical University of Denmark, Lyngby, Denmark}
\author{E.~Gschwendtner}
\affiliation{CERN, Geneva, Switzerland} 
\author{J.~Hansen}
\affiliation{CERN, Geneva, Switzerland} 
\author{A.~Helm}
\affiliation{GoLP/Instituto de Plasmas e Fus\~{a}o Nuclear, Instituto Superior T\'{e}cnico, Universidade de Lisboa, Lisbon, Portugal}
\author{J.R.~Henderson}
\affiliation{Lancaster University, Lancaster, UK}
\affiliation{Cockcroft Institute, Daresbury, UK}
\author{M.~H{\"u}ther}
\affiliation{Max Planck Institute for Physics, Munich, Germany}
\author{M.~Ibison}
\affiliation{University of Liverpool, Liverpool, UK}
\affiliation{Cockcroft Institute, Daresbury, UK}
\author{L.~Jensen}
\affiliation{CERN, Geneva, Switzerland}
\author{S.~Jolly}
\affiliation{UCL, London, UK}
\author{F.~Keeble}
\affiliation{UCL, London, UK}
\author{S.-Y.~Kim}
\affiliation{UNIST, Ulsan, Republic of Korea}
\author{F.~Kraus}
\affiliation{Philipps-Universit{\"a}t Marburg, Marburg, Germany}
\author{Y.~Li}
\affiliation{University of Manchester, Manchester, UK}
\affiliation{Cockcroft Institute, Daresbury, UK}
\author{S.~Liu}
\affiliation{TRIUMF, Vancouver, Canada}
\author{N.~Lopes}
\affiliation{GoLP/Instituto de Plasmas e Fus\~{a}o Nuclear, Instituto Superior T\'{e}cnico, Universidade de Lisboa, Lisbon, Portugal}
\author{K.V.~Lotov}
\affiliation{Budker Institute of Nuclear Physics SB RAS, Novosibirsk, Russia}
\affiliation{Novosibirsk State University, Novosibirsk, Russia}
\author{L.~Maricalva~Brun}
\affiliation{CERN, Geneva, Switzerland}
\author{M.~Martyanov}
\affiliation{Max Planck Institute for Physics, Munich, Germany}
\author{S.~Mazzoni}
\affiliation{CERN, Geneva, Switzerland}
\author{D.~Medina~Godoy}
\affiliation{CERN, Geneva, Switzerland}
\author{V.A.~Minakov}
\affiliation{Budker Institute of Nuclear Physics SB RAS, Novosibirsk, Russia}
\affiliation{Novosibirsk State University, Novosibirsk, Russia}
\author{J.~Mitchell}
\affiliation{Lancaster University, Lancaster, UK}
\affiliation{Cockcroft Institute, Daresbury, UK}
\author{J.C.~Molendijk}
\affiliation{CERN, Geneva, Switzerland}
\author{J.T.~Moody}
\affiliation{Max Planck Institute for Physics, Munich, Germany}
\author{M.~Moreira}
\affiliation{GoLP/Instituto de Plasmas e Fus\~{a}o Nuclear, Instituto Superior T\'{e}cnico, Universidade de Lisboa, Lisbon, Portugal}
\affiliation{CERN, Geneva, Switzerland}
\author{P.~Muggli}
\affiliation{Max Planck Institute for Physics, Munich, Germany}
\affiliation{CERN, Geneva, Switzerland} 
\author{E.~{\"O}z}
\affiliation{Max Planck Institute for Physics, Munich, Germany}
\author{C.~Pasquino}
\affiliation{CERN, Geneva, Switzerland} 
\author{A.~Pardons}
\affiliation{CERN, Geneva, Switzerland}
\author{F.~Pe\~na~Asmus}
\affiliation{Max Planck Institute for Physics, Munich, Germany}
\affiliation{Technical University Munich, Munich, Germany}
\author{K.~Pepitone}
\affiliation{CERN, Geneva, Switzerland}
\author{A.~Perera}
\affiliation{University of Liverpool, Liverpool, UK}
\affiliation{Cockcroft Institute, Daresbury, UK}
\author{A.~Petrenko}
\affiliation{CERN, Geneva, Switzerland}
\affiliation{Budker Institute of Nuclear Physics SB RAS, Novosibirsk, Russia}
\author{S.~Pitman}
\affiliation{Lancaster University, Lancaster, UK}
\affiliation{Cockcroft Institute, Daresbury, UK}
\author{A.~Pukhov}
\affiliation{Heinrich-Heine-University of D{\"u}sseldorf, D{\"u}sseldorf, Germany}
\author{S.~Rey}
\affiliation{CERN, Geneva, Switzerland}
\author{K.~Rieger}
\affiliation{Max Planck Institute for Physics, Munich, Germany}
\author{H.~Ruhl}
\affiliation{Ludwig-Maximilians-UniversitŠ\"{a}t, Munich, Germany}
\author{J.S.~Schmidt}
\affiliation{CERN, Geneva, Switzerland}
\author{I.A.~Shalimova}
\affiliation{Novosibirsk State University, Novosibirsk, Russia}
\affiliation{Institute of Computational Mathematics and Mathematical Geophysics SB RAS, Novosibirsk, Russia}
\author{P.~Sherwood}
\affiliation{UCL, London, UK}
\author{L.O.~Silva}
\affiliation{GoLP/Instituto de Plasmas e Fus\~{a}o Nuclear, Instituto Superior T\'{e}cnico, Universidade de Lisboa, Lisbon, Portugal}
\author{L.~Soby}
\affiliation{CERN, Geneva, Switzerland}
\author{A.P.~Sosedkin}
\affiliation{Budker Institute of Nuclear Physics SB RAS, Novosibirsk, Russia} 
\affiliation{Novosibirsk State University, Novosibirsk, Russia}
\author{R.~Speroni}
\affiliation{CERN, Geneva, Switzerland} 
\author{R.I.~Spitsyn}
\affiliation{Budker Institute of Nuclear Physics SB RAS, Novosibirsk, Russia}
\affiliation{Novosibirsk State University, Novosibirsk, Russia}
\author{P.V.~Tuev}
\affiliation{Budker Institute of Nuclear Physics SB RAS, Novosibirsk, Russia}
\affiliation{Novosibirsk State University, Novosibirsk, Russia}
\author{M.~Turner}
\affiliation{CERN, Geneva, Switzerland}
\author{F.~Velotti}
\affiliation{CERN, Geneva, Switzerland}
\author{L.~Verra}
\affiliation{CERN, Geneva, Switzerland}
\affiliation{University of Milan, Milan, Italy}
\author{V.A.~Verzilov}
\affiliation{TRIUMF, Vancouver, Canada} 
\author{J.~Vieira}
\affiliation{GoLP/Instituto de Plasmas e Fus\~{a}o Nuclear, Instituto Superior T\'{e}cnico, Universidade de Lisboa, Lisbon, Portugal}
\author{C.P.~Welsch}
\affiliation{University of Liverpool, Liverpool, UK}
\affiliation{Cockcroft Institute, Daresbury, UK}
\author{B.~Williamson}
\affiliation{University of Manchester, Manchester, UK}
\affiliation{Cockcroft Institute, Daresbury, UK}
\author{M.~Wing}
\thanks{Corresponding author.}
\affiliation{UCL, London, UK}
\author{B.~Woolley}
\affiliation{CERN, Geneva, Switzerland}
\author{G.~Xia}
\affiliation{University of Manchester, Manchester, UK}
\affiliation{Cockcroft Institute, Daresbury, UK}
\collaboration{The AWAKE Collaboration}
\noaffiliation

\date{\today}

\maketitle

\textbf{
High energy particle accelerators have been crucial in providing a deeper understanding of fundamental particles and the forces that govern their interactions.  In order to increase the energy or reduce the size of the accelerator, new acceleration schemes need to be developed.  Plasma wakefield acceleration~\cite{Tajima:1979bn,Chen:1984up,pop-14-055501,RevModPhys.81.1229,doi:10.1142/S1793626816300036}, in which the electrons in a plasma are excited, leading to strong electric fields, is one such promising novel acceleration technique.  Pioneering experiments have shown that an intense laser pulse~\cite{Modena:1995dgw,Mangles:2004ta,Geddes:2004tb,Faure:2004tc} or electron bunch~\cite{Blumenfeld:2007ph,Litos:2014} traversing a plasma, drives electric fields of tens of giga-volts per metre and above.  These values are well beyond those achieved in conventional radio frequency (RF) accelerators which are limited to about 0.1\,giga-volt per metre.  A limitation of laser pulses and electron bunches is their low stored energy, which motivates the use of multiple stages to reach very high energies~\cite{doi:10.1142/S1793626816300036,PhysRevSTAB.13.101301}. The use of proton bunches is compelling, as they have the potential to drive wakefields and accelerate electrons to high energy in a single accelerating stage~\cite{Caldwell:2008ak}.  The long proton bunches currently available can be used, as they undergo a process called self-modulation~\cite{PhysRevLett.104.255003,PhysRevLett.107.145002,PhysRevLett.107.145003}, a particle--plasma interaction which longitudinally splits the bunch into a series of high density microbunches, which then act resonantly to create large wakefields.  The Advanced Wakefield (AWAKE) experiment at CERN~\cite{Caldwell:2015rkk,Gschwendtner:2015rni,Muggli:2017rkx} uses intense bunches of protons, each of energy 400\,giga-electronvolts (GeV), with a total bunch energy of 19\,kilojoules, to drive a wakefield in a 10\,metre long plasma.  Bunches of electrons are injected into the wakefield formed by the proton microbunches.  This paper presents measurements of electrons accelerated up to 2\,GeV at the AWAKE experiment.  This constitutes the first demonstration of proton-driven plasma wakefield acceleration. The potential for this scheme to produce very high energy electron bunches in a single accelerating stage~\cite{doi:10.1063/1.3641973} means that the results shown here are a significant step towards the development of future high energy particle accelerators~\cite{Caldwell2016, Xia:2014}.
}

\par
\begin{figure*}[t]
		\includegraphics[width=\textwidth]{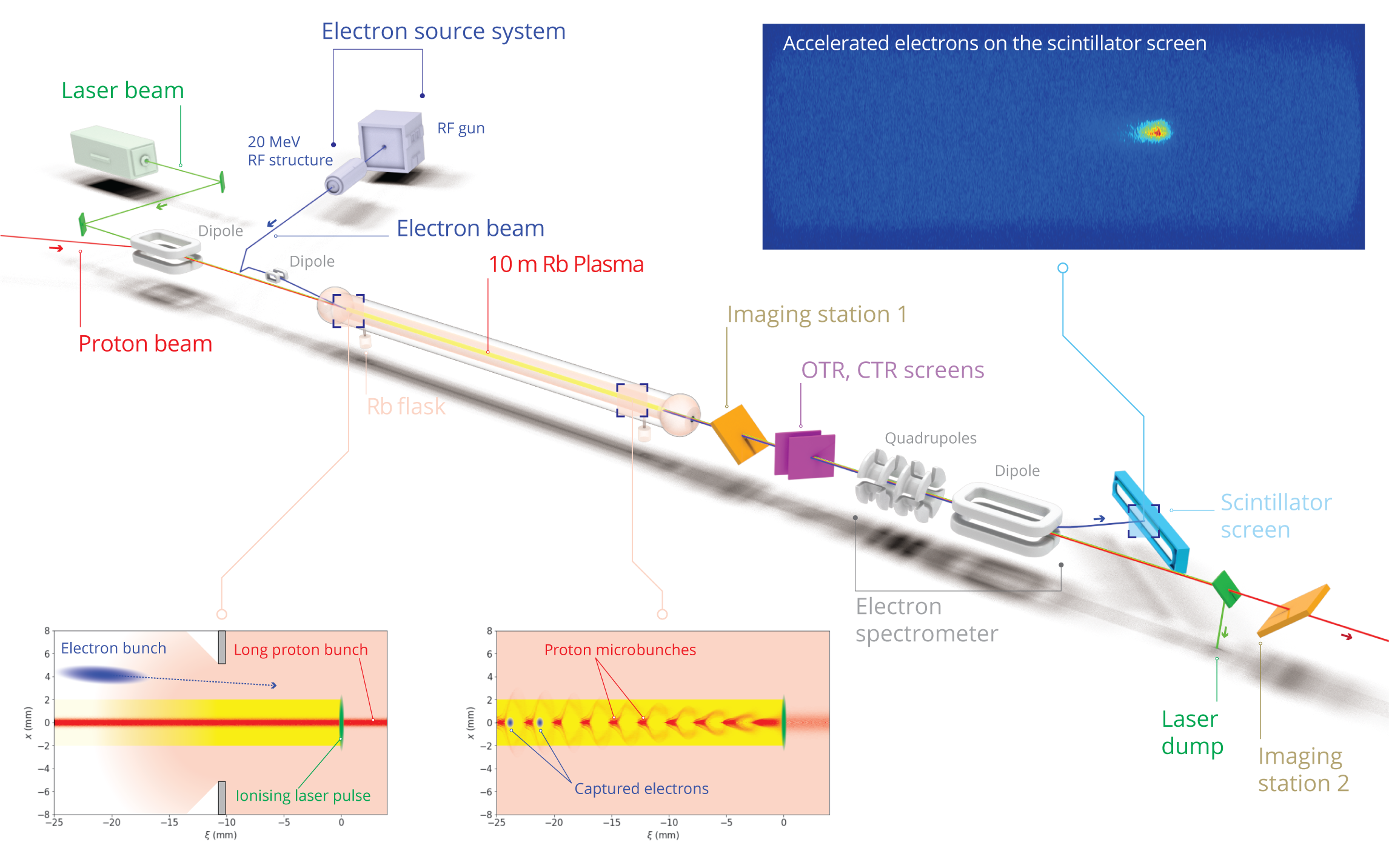}
		\caption{\label{fig:layout}The layout of the AWAKE experiment. The proton bunch and laser pulse propagate from left to right across the image, through a 10\,m column of rubidium vapour. This laser pulse (green, bottom images) singly ionises the rubidium (Rb) to form a plasma (yellow) which then interacts with the proton bunch (red, bottom left image). This interaction modulates the long proton bunch into a series of microbunches (bottom right image) which drive a strong wakefield in the plasma.  The self-modulation of the proton bunch is measured in imaging stations 1 and 2 and the optical and coherent transition radiation (OTR, CTR) diagnostics.  The rubidium is supplied by two flasks (pink) at each end of the vapour source. The density is controlled by changing the temperature in these flasks and a gradient may be introduced by changing their relative temperature. Electrons (blue), generated using a radio frequency (RF) source, propagate a short distance behind the laser pulse and are injected into the wakefield by crossing at an angle. Some of these electrons are captured in the wakefield and accelerated to high energies. The accelerated electron bunches are focused and separated from the protons by the spectrometer's quadrupoles and dipole magnet (grey, right). These electrons interact with a scintillating screen (top right image), allowing them to be imaged and their energy inferred from their position.}
\end{figure*}

The layout of the AWAKE experiment is shown in FIG.~\ref{fig:layout}.  A proton bunch from the CERN Super Proton Synchrotron (SPS) accelerator co-propagates with a laser pulse (green) which creates a plasma (yellow) in a column of rubidium vapour (pink) and seeds the modulation of the proton bunch into microbunches (FIG.~\ref{fig:layout}; red, bottom images). The protons have an energy of 400\,GeV and the root mean square (rms) bunch length is 6--8\,cm~\cite{Gschwendtner:2015rni}. The bunch is focused to a transverse size of approximately 200\,$\mu$m rms at the entrance of the vapour source, with the bunch population varying shot-to-shot in the range \mbox{$N_p \simeq2.5$--$3.1\times10^{11}$} protons per bunch. Proton extraction occurs every 15--30\,s. The laser pulse used to singly ionise the rubidium (Rb) in the vapour source~\cite{OZ2014197,GPlyushc2018} is 120 fs-long with a central wavelength of 780\,nm and a maximum energy of 450\,mJ~\cite{Fedosseev:2016ccm}.  The pulse is focused to a waist of approximately 1\,mm FWHM (full width at half maximum) inside the Rb vapour source, five times the transverse size of the proton bunch.  The Rb vapour source (FIG.~\ref{fig:layout}; centre) is of length 10\,m and diameter 4\,cm, with Rb flasks at each end. The Rb vapour density and, hence, the plasma density $n_{pe}$ can be varied in the range \mbox{$10^{14}$--$10^{15}\,\mathrm{cm}^{-3}$} by heating the Rb flasks to temperatures of 160--$210\,^{\circ}$C. This density range corresponds to a plasma wavelength of 1.1--3.3\,mm, as detailed in the Methods section. A plasma density gradient can be introduced by heating the Rb flasks to different temperatures. Heating the downstream (FIG.~\ref{fig:layout}; right side) flask to a higher temperature than the upstream (left side) flask creates a positive density gradient and vice versa. Plasma density gradients have been shown in simulation to produce significant increases in the maximum energy attainable by the injected electrons~\cite{PETRENKO201663}.  The effect of density gradients here is different from that for short drivers~\cite{PhysRevE.63.056405}.  In addition to keeping the wake travelling at the speed of light at the witness position, the gradient prevents destruction of the bunches at the final stage of self-modulation~\cite{doi:10.1063/1.4933129}, thus increasing the wakefield amplitude at the downstream part of the plasma cell.  The Rb vapour density is constantly monitored by an interferometer-based diagnostic~\cite{Batsch2018}.
\par
The self-modulation of the proton bunch into microbunches (FIG.~\ref{fig:layout}; red, bottom right image) is measured using optical and coherent transition radiation (OTR, CTR) diagnostics (FIG.~\ref{fig:layout}; purple)~\cite{Muggli:2015qwm}. However, these diagnostics have a destructive effect on the accelerated electron bunch and cannot be used during electron acceleration experiments. The second beam imaging station (FIG.~\ref{fig:layout}; orange, right) is used instead, providing an indirect measurement of the self-modulation by measuring the transversely defocused protons~\cite{TURNER2017100}. These protons are expelled from the central propagation axis by transverse electric fields that are only present when the proton bunch undergoes modulation in the plasma.
\par
Electron bunches with a charge of $656\pm14$\,pC are produced and accelerated to $18.84\pm0.05$\,MeV in an RF structure upstream of the vapour source~\cite{PEPITONE2018}.  These electrons are then transported along a beam line before being injected into the vapour source. Magnets along the beam line are used to control the injection angle and focal point of the electrons. For the results presented here, the electrons enter the plasma with a small vertical offset with respect to the proton bunch and a 200\,ps delay with respect to the ionising laser pulse (FIG.~\ref{fig:layout}, bottom left). The beams cross approximately 2\,m into the vapour source at a crossing angle of 1.2--2\,mrad. Simulations show that electrons are captured in larger numbers and accelerated to higher energies when injected off-axis rather than collinearly with the proton bunch~\cite{Caldwell:2015rkk}. The normalised emittance of the witness electron beam at injection is approximately 11--14\,mm\,mrad and its focal point is close to the entrance of the vapour source. The electron bunch delay of 200\,ps corresponds to approximately 25 proton microbunches resonantly driving the wakefield at $n_{pe}=2\times10^{14}\,\mathrm{cm}^{-3}$ and 50 microbunches at $n_{pe}=7\times10^{14}\,\mathrm{cm}^{-3}$.
\par
A magnetic electron spectrometer (FIG.~\ref{fig:layout}, right) allows measurement of the accelerated electron bunch~\cite{Deacon:2141860}. Two quadrupole magnets are located 4.48\,m and 4.98\,m downstream of the vapour source exit iris and focus the witness beam vertically and horizontally respectively, in order to more easily identify a signal. These are followed by a 1\,m long C-shaped electromagnetic dipole with a maximum magnetic field of approximately 1.4\,T. A large triangular vacuum chamber sits in the cavity of the dipole.  This chamber is designed to keep accelerated electron bunches under vacuum whilst the magnetic field of the dipole induces an energy-dependent horizontal deflection in the bunch. Electrons within a specific energy range then exit this vacuum chamber through a 2\,mm thick aluminium window and are incident on a 0.5\,mm thick gadolinium oxysulfide (Gd$_{2}$O$_{2}$S:Tb) scintillator screen (FIG.~\ref{fig:layout}; blue, right) attached to its exterior surface. The proton bunch is not significantly affected by the spectrometer magnets due to its considerably higher momentum and continues to the beam dump. The scintillating screen is 997\,mm wide and $62\,$mm high with semi-circular ends. Light emitted from the scintillator screen is transported over a distance of $17\,$m via three highly reflective optical-grade mirrors to an intensified charge-couple device (CCD) camera fitted with a $400\,$mm focal length lens. The camera and the final mirror of this optical line are housed in a dark room which reduces ambient light incident on the camera to negligible values.
\par
The energy of the accelerated electrons is inferred from their horizontal position in the plane of the scintillator. The relationship between this position and the electron's energy is dependent on the strength of the dipole, which can be varied from approximately 0.1--1.4\,T. This position--energy relationship has been simulated using the Beam Delivery Simulation (\textsc{BDSIM}) code~\cite{Nevay2018}. The simulation tracks electrons of various energies through the spectrometer using measured and simulated magnetic field maps for the spectrometer dipole, as well as the relevant distances between components. The accuracy of the magnetic field maps, the precision of the distance measurements and the 1.5\,mm resolution of the optical system lead to an energy uncertainty of approximately 2\%. The overall uncertainty, however, is dominated by the emittance of the accelerated electrons, and can be larger than 10\%. The use of the focusing quadrupoles limits this uncertainty to approximately 5\% for electrons near to the focused energy.
\par
Due to the difficulty of propagating an electron beam of well known intensity to the spectrometer at AWAKE, the charge response of the scintillator is calculated using data acquired at the CERN Linear Electron Accelerator for Research (CLEAR) facility. This calibration is performed by placing the scintillator and vacuum window next to a beam charge monitor on the CLEAR beam line and measuring the scintillator signal. The response of the scintillator is found to depend linearly on charge over the range 1--50\,pC. The response is also found to be independent of position and of energies in the range \mbox{100--180\,MeV}, to within the measurement uncertainty. This charge response is then recalculated for the spectrometer's optical system at AWAKE by imaging a well known light source at both locations. A response of $6.9\pm2.1\times10^{6}$ CCD counts per incident pC of charge, given the acquisition settings used at AWAKE, is determined. The large uncertainty is due to different triggering conditions at CLEAR and AWAKE and systematic uncertainties in the calibration results.
\par
\begin{figure}[t]
\includegraphics[width=\columnwidth]{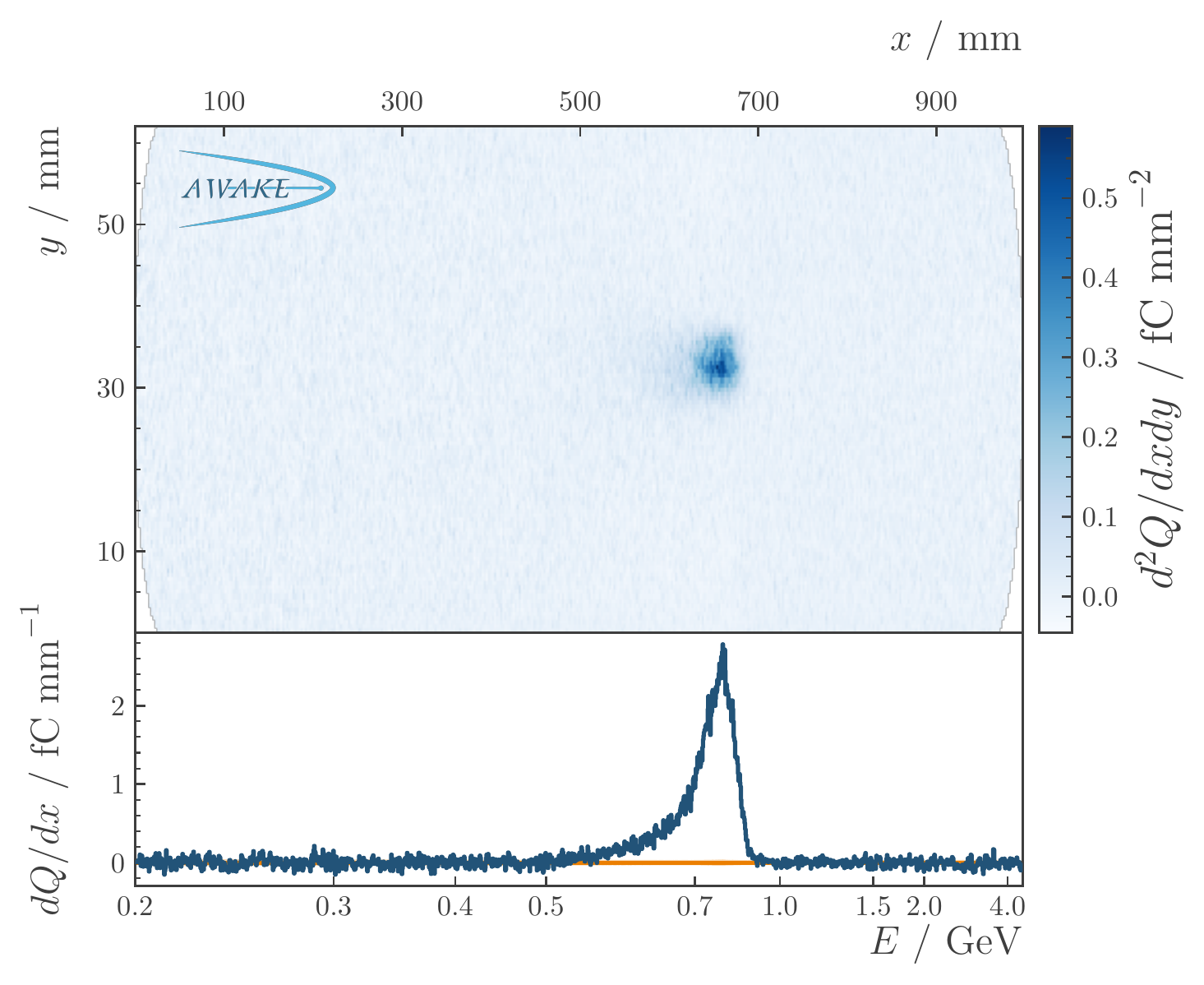}
\caption{\label{fig:im}Signal of accelerated electrons.  An image of the scintillator (horizontal distance, $x$, and vertical distance, $y$) with an electron signal clearly visible (top) and a vertical integration over the observed charge in the central region of the image (bottom), with background subtraction and geometric corrections applied, is shown. The intensity of the image is given in charge, $Q$, per unit area, calculated using the central value from the calibration of the scintillator. A $1\,\sigma$ uncertainty band from the background subtraction is shown in orange around zero on the bottom plot. Both the image and the projection are binned in space, as shown on the top axis, but the central value from the position--energy conversion is indicated at various points on the bottom axis. The electron signal is clearly visible above the noise, with a peak intensity at energy, $E \sim 800$\,MeV.}
\end{figure}
Reliable acceleration of electrons relies on reproducible self-modulation of the proton beam. As well as the observation of the transverse expansion of the proton bunch, the OTR and CTR diagnostics showed clear microbunching of the beam. The proton microbunches were observed to be separated by the plasma wavelength (inferred from the measured Rb vapour density, see Methods section) for all parameter ranges investigated; they were also reproducible and stable in phase relative to the seeding. The detailed study of the self-modulation process will be the subject of separate AWAKE publications.
\par
The data presented here were taken in May 2018.  The top of FIG.~\ref{fig:im} shows an image of the scintillator from an electron acceleration event at a plasma density of $1.8\times10^{14}\,\mathrm{cm}^{-3}$ with a $+5.3\%\pm0.3\%$ density difference over 10\,m, in the direction of the proton bunch propagation. This image has been background-subtracted and corrected for vignetting and electron angle effects, as described in the Methods section. The spectrometer's quadrupoles were focusing at an energy of approximately 700\,MeV during this event, creating a significant reduction in the vertical spread of the beam. Below the image is a projection obtained by integrating over a central region of the scintillator. A $1\,\sigma$ uncertainty band coming from the background subtraction is shown around zero. The peak in this figure has a high signal-to-noise ratio, giving clear evidence of accelerated electrons. In both the image and the projection, the charge density is calculated using the central value of $6.9\times10^{6}$ CCD counts per pC. The asymmetric shape of the peak is due to the nonlinear position--energy relationship induced in the electron bunch by the magnetic field; when re-binned in energy, the signal peak is approximately Gaussian. Accounting for the systematic uncertainties described earlier, the observed peak has a mean of $800\pm40\,\mathrm{MeV}$, a FWHM of $137.3\pm13.7\,\mathrm{MeV}$ and a total charge of $0.249\pm0.074\,\mathrm{pC}$. The amount of charge captured is expected to increase considerably~\cite{Caldwell:2015rkk} as the emittance of the injected electron bunch is reduced and its geometrical overlap with the wakefield is improved.
\par
\begin{figure}[t]
\includegraphics[width=\columnwidth]{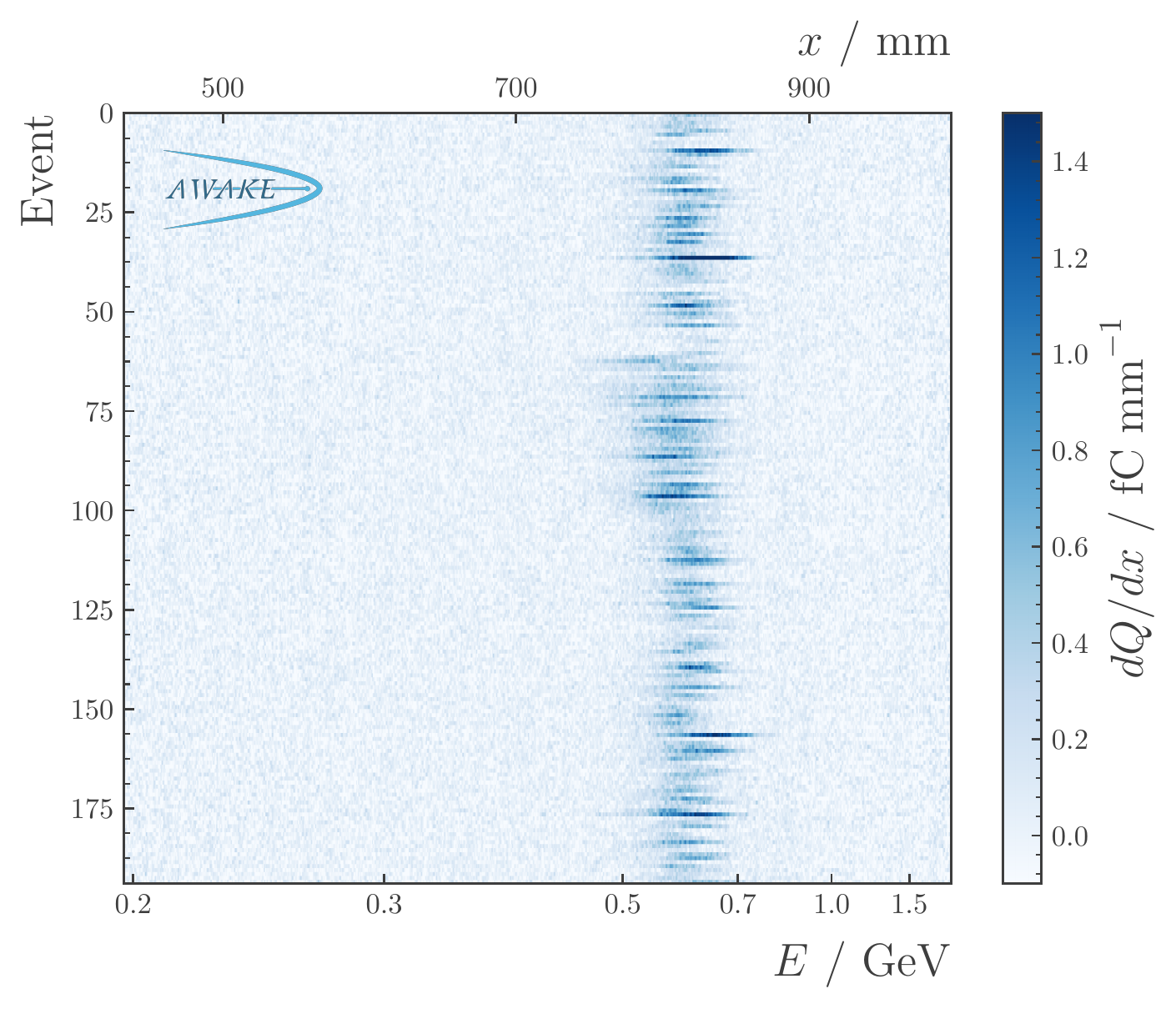}
\caption{\label{fig:water}Background-subtracted projections of consecutive electron-injection events. Each projection is a vertical integration over the central region of a background-subtracted spectrometer camera image. Brighter colours indicate regions of high charge density, $dQ/dx$, corresponding to accelerated electrons. The spectrometer's quadrupoles were varied to focus at energies of 460--620\,MeV over the duration of the dataset. No other parameters were deliberately varied. The consistent peak around energy $E \sim 600$\,MeV demonstrates the stability and reliability of the electron acceleration.}
\end{figure}
The stability and reliability of the electron acceleration is evidenced by FIG.~\ref{fig:water}, which shows projections from many electron-injection consecutive events. Each row in this plot is the background subtracted projection from a single event, with colour representing the signal intensity. The events correspond to a two hour running period during which the quadrupoles were varied to focus over a range of approximately 460--620\,MeV. Other parameters, such as the proton bunch population were not deliberately changed but naturally vary on a shot-to-shot basis. Despite the quadrupole scan and the natural fluctuations in the beam parameters, the plot still shows consistent and reproducible acceleration of electron bunches to approximately 600\,MeV. The plasma density for these events is $1.8\times10^{14}\,\mathrm{cm}^{-3}$, with no density gradient. This lack of gradient is the cause of the difference in energy between the event in FIG.~\ref{fig:im} and the events in FIG.~\ref{fig:water}.
\par
\begin{figure}[t]
\includegraphics[width=\columnwidth]{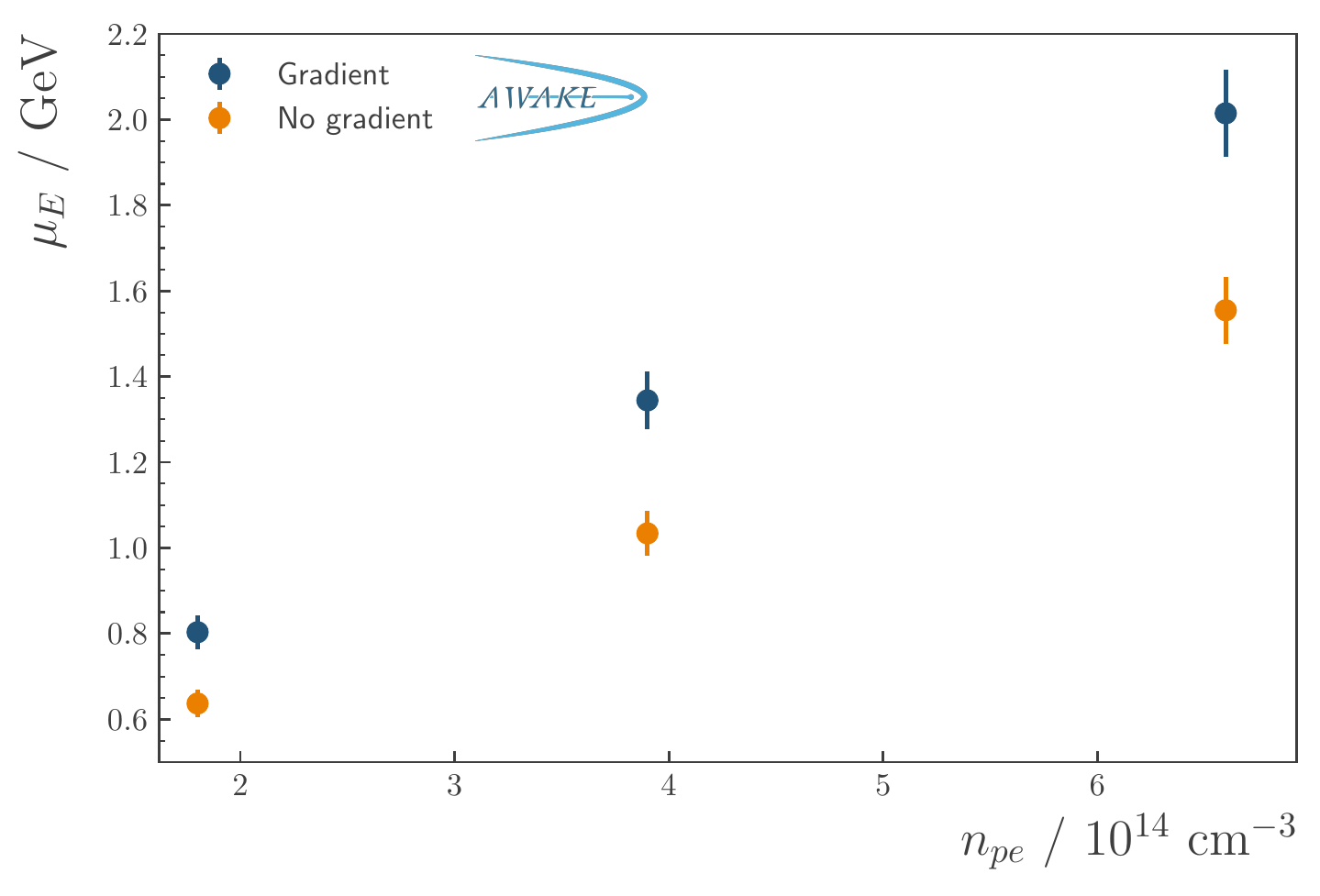}
\caption{\label{fig:energy}Measurement of the highest peak energies, $\mu_E$, achieved at different plasma densities, $n_{pe}$, with and without plasma density gradients. The gradients chosen are those which are observed to maximise the energy gain. Acceleration to $2.0\pm0.1$\,GeV is achieved with a plasma density of \mbox{$6.6\times10^{14}\,\mathrm{cm}^{-3}$} with a $+2.2\%\pm0.1\%$ plasma density difference over 10\,m.}
\end{figure}
The energy gain achievable by introducing a more optimal gradient is demonstrated in FIG.~\ref{fig:energy}, which shows the peak energy achieved at different plasma densities with and without a gradient. The density gradients chosen are those that are observed to maximise the peak energy for a given plasma density.  At \mbox{$1.8\times10^{14}\,\mathrm{cm}^{-3}$} the density difference was approximately $+5.3\%\pm0.3\%$ over 10\,m, while at \mbox{$3.9\times10^{14}\,\mathrm{cm}^{-3}$} and \mbox{$6.6\times10^{14}\,\mathrm{cm}^{-3}$} it fell to $+2.5\%\pm0.3\%$ and $+2.2\%\pm0.1\%$, respectively.  Given the precise control of the longitudinal plasma density, the small values of density gradient can have a significant effect on the acceleration where the electrons are injected many 10s of microbunches behind the ionising laser pulse~\cite{PETRENKO201663}.  The charge of the observed electron bunches decreases at higher plasma densities due, in part, to the smaller transverse size of the wakefield. Additionally, the spectrometer's quadrupoles have a maximum focusing energy of 1.3\,GeV making bunches accelerated to higher energies than this harder to detect above the background noise.
The energies shown in FIG.~\ref{fig:energy} are determined by binning the pixel data in energy and fitting a Gaussian over the electron signal region; the peak energy $\mu_{E}$ is the mean of this Gaussian.  The observed energy spread of each bunch is determined by the width of this Gaussian and is approximately 10\% of the peak energy.
The peak energy increases with the density, reaching $2.0\pm0.1$\,GeV for \mbox{$n_{pe}=6.6\times10^{14}\,\mathrm{cm}^{-3}$} with a density gradient, at which point the charge capture is significantly lower. The energies of the accelerated electrons are within the range of values originally predicted by particle-in-cell and fluid code simulations of the AWAKE experiment~\cite{Caldwell:2015rkk,Gschwendtner:2015rni,PETRENKO201663}. Future data taking runs will address the effect of the electron bunch delay, injection angle and other parameters on accelerated energy and charge capture. These studies will help determine what sets the limit on the energy gain.
\par
In summary, proton-driven plasma wakefield acceleration has been demonstrated for the first time.  The strong electric fields, generated by a series of proton microbunches, were sampled with a bunch of electrons. These electrons were accelerated up to 2\,GeV in approximately 10\,m of plasma and measured using a magnetic spectrometer.  This technique has the potential to accelerate electrons to the TeV scale in a single accelerating stage.  Although still in the early stages of its programme, the AWAKE collaboration has taken an important step on the way to realising new high energy particle physics experiments.
\hfill \break
\begin{acknowledgments}
This work was supported in parts by: a Leverhulme Trust Research Project Grant RPG-2017-143 and by STFC (AWAKE-UK, Cockroft Institute core and UCL consolidated grants), United Kingdom; the Russian Science Foundation (project No.\ 14-50-00080) for simulations of oblique injection performed by Budker INP group; a Deutsche Forschungsgemeinschaft project grant PU 213-6/1 ``Three-dimensional quasi-static simulations of beam self-modulation for plasma wakefield acceleration''; the National Research Foundation of Korea (Nos.\ NRF-2015R1D1A1A01061074 and NRF-2016R1A5A1013277); the Portuguese FCT---Foundation for Science and Technology, through grants CERN/FIS-TEC/0032/2017, PTDC-FIS-PLA-2940-2014, UID/FIS/50010/2013 and SFRH/IF/01635/2015; NSERC and CNRC for TRIUMF's contribution; and the Research Council of Norway. M. Wing acknowledges the support of the Alexander von Humboldt Stiftung and DESY, Hamburg.  For their advice and contributions to the development of the magnetic spectrometer, we gratefully acknowledge B.~Biskup, P. La~Penna (European Southern Observatory (ESO)) and M. Quattri (ESO). A. Petrenko acknowledges G. Demeter  (Wigner Institute, Budapest) for the calculation of rubidium ionisation probability at AWAKE. F. Keeble acknowledges the operators of the CLEAR facility for their assistance during the calibration of the spectrometer. The AWAKE collaboration acknowledge the SPS team for their excellent proton delivery.
\end{acknowledgments}

\appendix*
\section{Methods}
\subsection{Plasma generation}
A CentAurus Ti:Sapphire laser system is used to ionise the Rb in the vapour source. The Rb is confined by expansion chambers at the ends of the source with 10\,mm diameter irises through which Rb constantly flows and condensates on the expansion walls. By the relation $\lambda_{pe} = 2 \pi c \sqrt{\epsilon_0 m_e/n_{pe} e^2}$, where $c$ is the speed of light, $\epsilon_0$ is the permittivity of free space, $m_e$ is the electron mass and $e$ is the electron charge, the available density range of \mbox{$10^{14}$--$10^{15}\,\mathrm{cm}^{-3}$} corresponds to a plasma wavelength of \mbox{$\lambda_{pe}\simeq1.1$--$3.3\,$mm}. The vapour density uniformity is ensured by flowing a heat exchanging fluid around a concentric tube surrounding the source at a temperature stabilised to $\pm0.05\,^{\circ}$C. Longitudinal density differences between $-10\%$ and $+10\%$ over 10\,m, may be implemented and can be controlled at the percent-level.  It is noted that the motion of the Rb ions can be neglected during the transit of the proton bunch because the heavy Rb ions are singly ionised~\cite{PhysRevLett.109.145005}.
\subsection{Witness electron beam}
Production of the witness electron beam is initiated by the illumination of a Cs$_{2}$Te cathode by a frequency-tripled laser pulse derived from the ionising laser. Electron bunches with a charge of $656\pm14$\,pC are produced and accelerated to an energy of 5.5\,MeV in a 2.5 cell RF-gun and are subsequently accelerated up to $18.84\pm0.05$\,MeV using a 30 cell travelling wave structure.  These electrons are then transported along an 18\,m beam line before being injected into the vapour source. The focal point and crossing angle of the witness beam can be controlled via a combination of quadrupole and kicker magnets along this beam line.
\subsection{Background subtraction}
The large distance between the camera and the proton beam line means that background noise generated by radiation directly incident on the CCD is minimal. The spectrometer's scintillator, however, is subject to significant background radiation. The rise and decay of the scintillator signal occur on timescales longer than $1\,\mu\mathrm{s}$ and, as such, the scintillator photons captured by the camera are produced by an indivisible combination of background radiation and accelerated electrons. The majority of this background radiation is due to the passage of the proton bunch and comes from two main sources: a 0.2\,mm thick aluminium window located 43\,m upstream of the spectrometer between AWAKE and the SPS transfer line and a 0.6~mm thick aluminium iris at the downstream end of the vapour source. The inner radius of this iris is 5~mm, leading to negligible interaction with the standard SPS proton bunch. However, protons defocused during self-modulation, such as those measured at the downstream imaging station, can interact with the iris, creating a significant background. The strength of the transverse fields in the plasma and, hence, the number of protons defocused, is strongly dependent on the plasma density. Consequently, the background generated by the defocused protons is more significant at higher plasma densities, such as the AWAKE baseline density of $7\times10^{14}\,\mathrm{cm}^{-3}$. At this density, the radiative flux on the scintillator due to the iris is significantly higher than that from the thin window. Conversely, at a lower plasma density, such as $2\times10^{14}\,\mathrm{cm}^{-3}$, the radiation from the iris disappears completely and the remaining incident radiation is produced almost entirely by the interaction of the protons with the upstream window.
\par
Due to the variable nature of the radiation incident on the scintillator, background subtraction is a multistep process. A background data sample with the electron beam off at a plasma density of $1.8\times10^{14}\,\mathrm{cm}^{-3}$ is taken, such that the background has two key components: one due to the camera readout and ambient light in the experimental area and another $N_{p}$-dependent background caused by the proton bunch passing through the thin window. For each pixel imaging the scintillator, a linear function of $N_{p}$ is defined by a $\chi^{2}$ minimisation fit to the background data sample, giving an $N_{p}$-dependent mean background image. For each signal event, a region of the scintillator is chosen where no accelerated electrons are expected, typically the lowest energy part, and the background is rescaled by the ratio of the sums over this region in the signal event and the $N_{p}$-scaled background image. At higher plasma densities, a further step is needed to subtract the background from the iris. This background falls rapidly with increasing distance from the beam line and thus is dependent on the horizontal position in the plane of the scintillator. A new region where the expected number of accelerated electrons is small is chosen, this time along the top and bottom edges of the scintillator. The mean of each column of pixels in this region is calculated and then subtracted from each pixel in the central region of that same column, leaving only the signal. The semi-circular ends of the scintillator reduce the effectiveness of this technique at the highest and lowest energies.
\subsection{Signal extraction}
To give an accurate estimate of the electron bunch charge the background subtracted signal is corrected for two effects which vary across the horizontal plane of the scintillator. One effect comes from the variation in the electron's horizontal angle of incidence on the scintillator. This angle is determined by the same tracking simulation used to define the position--energy relationship and introduces a cosine correction to the signal due to the variation in the electron's path length through the scintillator. The second effect is vignetting which occurs due to the finite size of the spectrometer's optics and the angular emission profile of the scintillator photons. A lamp which mimics this emission profile is scanned across the horizontal plane of the scintillator and the vignetting correction is determined by measuring its relative brightness. The increase in radiation accompanying the electron bunch, due to its longer path length through the vacuum window at larger incident angles, is negligible and therefore does not require an additional correction factor.
\subsection{Data and code availability}
The datasets generated during and/or analysed during the current study are available from the corresponding author on reasonable request.  The software code used in the analysis and to produce Figs. 2, 3 and 4 are available from the corresponding author on reasonable request.

\end{document}